# Further Results on Quadratic Permutation Polynomial-Based Interleavers for Turbo Codes

Eirik Rosnes, *Senior Member, IEEE*

*Abstract*—An interleaver is a critical component for the channel coding performance of turbo codes. Algebraic constructions are of particular interest because they admit analytical designs and simple, practical hardware implementation. Also, the recently proposed quadratic permutation polynomial (QPP) based interleavers by Sun and Takeshita (*IEEE Trans. Inf. Theory*, Jan. 2005) provide excellent performance for short-to-medium block lengths, and have been selected for the 3GPP LTE standard. In this work, we derive some upper bounds on the best achievable minimum distance $d_{\min}$ of QPP-based conventional binary turbo codes (with tailbiting termination, or dual termination when the interleaver length $N$ is sufficiently large) that are tight for larger block sizes. In particular, we show that the minimum distance is at most $2(2^{\nu+1}+9)$, independent of the interleaver length, when the QPP has a QPP inverse, where $\nu$ is the degree of the primitive feedback and monic feedforward polynomials. However, allowing the QPP to have a larger degree inverse may give strictly larger minimum distances (and lower multiplicities). In particular, we provide several QPPs with an inverse degree of at least three for some of the 3GPP LTE interleaver lengths giving a $d_{\min}$ with the 3GPP LTE constituent encoders which is strictly larger than $50$. For instance, we have found a QPP for $N = 6016$ which gives an estimated $d_{\min}$ of $57$. Furthermore, we provide the *exact* minimum distance and the corresponding multiplicity for all 3GPP LTE turbo codes (with dual termination) which shows that the best minimum distance is $51$. Finally, we compute the best achievable minimum distance with QPP interleavers for all 3GPP LTE interleaver lengths $N \leq 4096$, and compare the minimum distance with the one we get when using the 3GPP LTE polynomials.

## I. Introduction

Turbo codes have gained considerable attention since their introduction by Berrou *et al.* in 1993 [1] due to their near-capacity performance and low decoding complexity. The conventional turbo code is a parallel concatenation of two identical recursive systematic convolutional encoders separated by a pseudo-random interleaver.

Interleavers for conventional binary turbo codes [2–10] have been extensively investigated. The dithered relative prime (DRP) interleavers [9, 10] and the almost regular permutation interleavers [4] are considered among the best classes of interleavers. Recently, Sun and Takeshita [2] suggested the use of permutation polynomial (PP) based interleavers over integer rings. In particular, quadratic polynomials were emphasized. A useful property of this class of interleavers is that they are fully algebraic and admit the computation of analytical upper bounds on the best achievable minimum distance $d_{\min}$. The use of higher degree PPs was recently considered by Takeshita in [11], and a simple coefficient test for cubic PPs was provided in [12].

The decoding of turbo codes is performed by an iterative process in which the so-called *extrinsic information* is exchanged between sub-blocks of the iterative decoder. There are typically two (or more) sub-blocks in an iterative turbo decoder, each implementing a soft-input soft-output decoding algorithm of a convolutional code. The parallel processing of iterative decoding of turbo codes is of interest for high-speed decoders, and this requires that the interleaver satisfies additional properties. In particular, to avoid memory contentions in parallelized decoding, the interleaver should be *contention-free* [4, 13, 14]. In this regard, algebraically constructed interleavers have some advantages compared to random constructions. For instance, in [3], Takeshita showed that all PPs generate maximum contention-free interleavers, i.e., every factor of the interleaver length becomes a possible degree of parallel processing of the decoder. Thus, this class of interleavers is very interesting from an implementation point of view, and quadratic permutation polynomial (QPP) interleavers were indeed recently selected for the turbo codes in the 3GPP LTE standard [15].

This work is a continuation of [16], in which the $d_{\min}$ of conventional binary turbo codes with QPP interleavers was considered in detail. In particular, in [16], large tables of optimum (in terms of the induced $d_{\min}$ and its corresponding multiplicity) QPPs for conventional binary turbo codes with 8-state and 16-state constituent encoders were presented for short interleaver lengths. In this work, however, we consider the minimum distance properties with longer QPP interleavers. In particular, we present several upper bounds on the best possible $d_{\min}$ with QPP interleavers and compute the *exact* minimum distance of all 3GPP LTE turbo codes (with dual termination [17]) for all possible lengths. Estimating the minimum distance of 3GPP LTE turbo codes has recently been addressed by several authors, see, for instance, [18, 19]. However, the *exact* minimum distance has only been computed for short interleaver lengths $N \leq 104$ [19].

The remainder of this paper is organized as follows. QPPs over integer rings and some of their properties are reviewed in Section II. Section III provides upper bounds on the optimum minimum distance for QPP-based turbo codes. A table providing the exact minimum distance and its corresponding multiplicity for all 3GPP LTE turbo codes (with dual termination) is presented in Section IV. In Section V, we present the results of an exhaustive computer search over the entire class of QPP interleavers. In particular, the best achievable minimum distance for all 3GPP LTE interleaver lengths $N \leq 4096$ and for several selected larger interleaver

This work was supported by the Research Council of Norway (NFR) under Grants 174982 and 183316.

E. Rosnes is with the Selmer Center, Department of Informatics, University of Bergen, N-5020 Bergen, Norway (e-mail: eirik@ii.uib.no).

lengths, is provided. The results are compared to those with the 3GPP LTE polynomials. Furthermore, we provide a table of several improved QPPs (compared to the ones selected for the 3GPP LTE standard) for medium-to-large interleaver lengths. Conclusions and a discussion of future work are given in Section VI.

## II. QPPs Over Integer Rings

In this section, we establish notation and restate the criterion for existence of QPPs over integer rings. The interested reader is referred to [2, 20] for further details. Given an integer $N \geq 2$, a polynomial $f(x) = f_1 x + f_2 x^2 \pmod{N}$, where $f_1$ and $f_2$ are nonnegative integers, is said to be a QPP over the ring of integers $\mathbb{Z}_N$ when $f(x)$ permutes $\{0, 1, 2, \ldots, N-1\}$ [2].

In this paper, let the set of primes be $\mathcal{P} = \{2, 3, 5, 7, \ldots\}$. Then an integer $N$ can be factored as $N = \prod_{p \in \mathcal{P}} p^{n_{N,p}}$, where $n_{N,p} \geq 1$ for a finite number of $p$'s and $n_{N,p} = 0$ otherwise. For example, if $N = 3888 = 2^4 \times 3^5$ we have $n_{3888,2} = 4$ and $n_{3888,3} = 5$. For a quadratic polynomial $f(x) = f_1 x + f_2 x^2 \pmod{N}$, we will abuse the previous notation by writing $f_2 = \prod_{p \in \mathcal{P}} p^{n_{F,p}}$, i.e., the exponents of the prime factors of $f_2$ will be written as $n_{F,p}$ instead of the more cumbersome $n_{f_2,p}$ because we will be mainly interested in the factorization of $f_2$.

Let us denote $a$ divides $b$ by $a|b$ and by $a \nmid b$ otherwise. The greatest common divisor of $a$ and $b$ is denoted by $\gcd(a, b)$. The necessary and sufficient condition for a quadratic polynomial $f(x)$ to be a PP is given in the following proposition.

*Proposition 1 ([2, 20]):* Let $N = \prod_{p \in \mathcal{P}} p^{n_{N,p}}$. The necessary and sufficient condition for a quadratic polynomial $f(x) = f_1 x + f_2 x^2 \pmod{N}$ to be a PP can be divided into two cases.
1) Either $2 \nmid N$ or $4|N$ (i.e., $n_{N,2} \neq 1$)
   $\gcd(f_1, N) = 1$ and $f_2 = \prod_{p \in \mathcal{P}} p^{n_{F,p}}, n_{F,p} \geq 1, \forall p$ such that $n_{N,p} \geq 1$.
2) $2|N$ and $4 \nmid N$ (i.e., $n_{N,2} = 1$)
   $f_1 + f_2$ is odd, $\gcd(f_1, \frac{N}{2}) = 1$, and $f_2 = \prod_{p \in \mathcal{P}} p^{n_{F,p}}, n_{F,p} \geq 1, \forall p$ such that $p \neq 2$ and $n_{N,p} \geq 1$.

For example, if $N = 256$, then we determine from case 1) of Proposition 1 that $f_1 \in \{1, 3, 5, \ldots, 255\}$ (set of numbers relatively prime to $N$) and $f_2 \in \{2, 4, 6, \ldots, 254\}$ (set of numbers that contains 2 as a factor). This gives us $128 \times 127 = 16256$ possible pairs of coefficients $f_1$ and $f_2$ that make $f(x)$ a PP.

It is shown in [11] that some QPPs $f(x) = f_1 x + f_2 x^2 \pmod{N}$ such that $f_2 \not\equiv 0 \pmod{N}$ degenerate to a linear permutation polynomial (LPP), i.e., there is an LPP equivalent to the QPP generating the same permutation over the integer ring $\mathbb{Z}_N$. QPPs that do not degenerate to an LPP are called *irreducible* QPPs. The following proposition from [11] can be used to check if a QPP is irreducible.

*Proposition 2 ([11]):* A QPP $f(x) = f_1 x + f_2 x^2 \pmod{N}$ is irreducible if and only if $N/\gcd(2f_2, N) \neq 1$.

Combining Propositions 1 and 2 we conclude that if the length $N$ can written as the product of distinct prime numbers, then either there are no QPPs for this length or all QPPs are equivalent to LPPs.

*Proposition 3 ([21]):* Let $N = \prod_{p \in \mathcal{P}} p^{n_{N,p}} \leq 2^{50}$, $\phi(k) = \prod_{l=k}^{2k-2} l$, and $f(x)$ a PP. Decompose $\phi(k)$ into prime factors and denote the exponent of the prime factor $p$ as $n_{\phi(k),p}$. Then, $f(x)$ has $\prod_{k=1}^{L} \gcd(k!, N)$ inverse polynomials with the least degree $L$ if and only if there is a smallest integer $L$ such that

$$n_{F,p} \geq \begin{cases} \max\left(\left\lceil \frac{n_{N,2} - n_{\phi(L+1),2}}{L} \right\rceil, 1\right), & \text{if } n_{N,2} > 1 \text{ and } p = 2 \\ 0, & \text{if } n_{N,2} \leq 1 \text{ and } p = 2 \\ \max\left(\left\lceil \frac{n_{N,p} - n_{\phi(L+1),p}}{L} \right\rceil, 1\right), & \text{if } n_{N,p} > 0 \text{ and } p \neq 2 \\ 0, & \text{if } n_{N,p} = 0 \text{ and } p \neq 2. \end{cases}$$

*Example 1:* Let $N = 5504 = 2^7 \times 43$. With $L = 3$ in Proposition 3, $\phi(L+1) = \phi(4) = 120 = 2^3 \times 3 \times 5$, and we get

$$n_{F,p} \geq \begin{cases} 2, & \text{if } p = 2 \\ 1, & \text{if } p = 43 \\ 0, & \text{otherwise.} \end{cases} \quad (1)$$

## III. Upper Bounds on the Optimum $d_{\min}$

In this section, we report several upper bounds on the optimum $d_{\min}$ for QPP-based conventional binary turbo codes of nominal rate $1/3$. These upper bounds can be used in the selection of good interleaver lengths with QPP interleavers, and also to efficiently reject bad QPP candidates in a computer search.

In general, let $f(x) = f_1 x + f_2 x^2 \pmod{N}$ denote a QPP and $g(x) = g_1 x + \cdots + g_L x^L$ an inverse polynomial of degree $L$. We assume tailbiting termination of the upper and lower constituent encoders throughout this section, but we remark that the derived upper bounds can be shown to hold with dual termination as well, i.e., the upper and lower constituent encoders are forced to begin and end in the zero state, when $N$ is sufficiently large. Details are provided at the end of the section. We start by deriving some partial upper bounds (i.e., upper bounds that apply for specific interleaver lengths $N$) with no constraints on an inverse polynomial.

### A. Partial Upper Bounds on the $d_{\min}$

In this section, we derive partial upper bounds (i.e., upper bounds that apply for some values of the interleaver length $N$) on the $d_{\min}$ that holds for any QPP, i.e., there are no constraints on the degree of an inverse polynomial.

*Theorem 1:* The minimum distance of a conventional binary turbo code (of nominal rate $1/3$) with feedback polynomial $1 + D^2 + D^3$ and feedforward polynomial $1 + D + D^3$, is upper-bounded by $38 + 12l$, where $l$ is some nonnegative integer, for all interleaver lengths $N$ that satisfy

$$n_{N,p} \leq \begin{cases} l + 4, & \text{if } p = 2 \\ 2, & \text{if } p = 7 \\ 1, & \text{otherwise} \end{cases}$$

when using QPP interleavers.

*Proof:* In Fig. 1, an input-weight 6 codeword is shown. The upper constituent codeword contains 3 input-weight 2 fundamental paths, while the lower constituent codeword contains

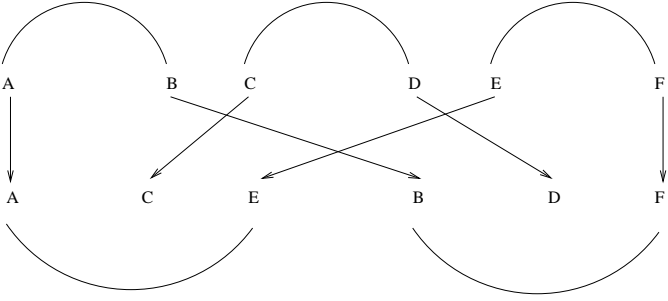

Fig. 1. An input-weight 6 critical codeword for QPP-based interleavers.

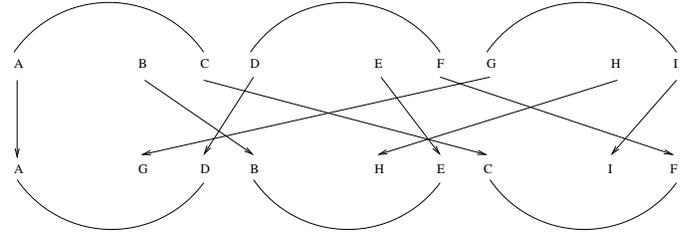

Fig. 2. An input-weight 9 critical codeword for QPP-based interleavers.

2 input-weight 3 fundamental paths. The interleaving of the systematic 1-positions is indicated by arrows and the 6 first letters of the English alphabet. The actual pattern in Fig. 1 is a codeword if

$$f(g(x+b)+a) \equiv f(g(x)+a)+b \pmod{N} \quad (2)$$
$$f(g(x+c)+a) \equiv f(g(x)+a)+c \pmod{N} \quad (3)$$

where $x \in \mathbb{Z}_N$ denotes the first systematic 1-position (labeled by "A" in Fig. 1) in the lower constituent codeword. For the input-weight 3 fundamental paths in the lower constituent codeword, the index differences between the second and first systematic 1-positions and the third and first systematic 1-positions are denoted by $b$ and $c$, respectively. For the input-weight 2 fundamental paths in the upper constituent codeword, the index difference between the final and first systematic 1-positions is denoted by $a$. The two congruences in (2) and (3) are equivalent, with $x = 0$, to

$$2baf_2(g_1 + bg_2 + \cdots + b^{L-1}g_L) \equiv 0 \pmod{N} \quad (4)$$
$$2caf_2(g_1 + cg_2 + \cdots + c^{L-1}g_L) \equiv 0 \pmod{N} \quad (5)$$

where $L$ is the degree of an inverse polynomial. These two congruences are satisfied if both $2baf_2 \equiv 0 \pmod{N}$ and $2caf_2 \equiv 0 \pmod{N}$. Choose $a = 2^l \cdot 7$, where $l$ is some nonnegative integer, and $b = 8$ and $c = 12$, from which it follows that the two congruences reduce to $2^{l+4} \cdot 7 \cdot f_2 \equiv 0 \pmod{N}$ and $2^{l+3} \cdot 21 \cdot f_2 \equiv 0 \pmod{N}$, and the result follows from Proposition 1 and the fact that the Hamming weight of the codeword in Fig. 1 is (at most) $38 + 12l$. ∎

*Theorem 2:* The minimum distance of a conventional binary turbo code (of nominal rate $1/3$) with feedback polynomial $1+D^2+D^3$ and feedforward polynomial and $1+D+D^3$, is upper-bounded by 51 for all interleaver lengths $N$ that satisfy

$$n_{N,p} \leq \begin{cases} 6, & \text{if } p = 2 \\ 2, & \text{if } p = 3 \\ 1, & \text{otherwise} \end{cases}$$

when using QPP interleavers.

*Proof:* In Fig. 2, an input-weight 9 codeword is shown. Both the upper and lower constituent codewords contain 3 input-weight 3 fundamental paths. The interleaving of the systematic 1-positions is indicated by arrows and the 9 first letters of the English alphabet. The actual pattern in Fig. 2 is

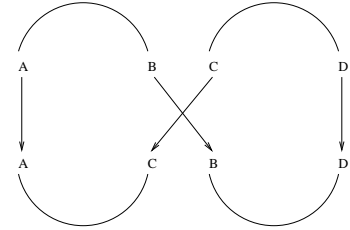

Fig. 3. An input-weight 4 critical codeword for QPP-based interleavers.

a codeword if

$$f(g(x)+b)+b \equiv f(g(x+b)+b) \pmod{N} \quad (6)$$
$$f(g(x)+b)+c \equiv f(g(x+c)+b) \pmod{N} \quad (7)$$
$$f(g(x)+c)+b \equiv f(g(x+b)+c) \pmod{N} \quad (8)$$
$$f(g(x)+c)+c \equiv f(g(x+c)+c) \pmod{N} \quad (9)$$

where $x \in \mathbb{Z}_N$ denotes the first systematic 1-position (labeled by "A" in Fig. 2) of the first fundamental path in the lower constituent codeword. The index differences between the second and first and the final and first systematic 1-positions of all the fundamental paths in both the upper and lower constituent codewords are denoted by $b$ and $c$, respectively. The congruences in (6) to (9) are equivalent, with $x = 0$, to

$$2b^2 f_2(g_1 + bg_2 + \cdots + b^{L-1}g_L) \equiv 0 \pmod{N}$$
$$2bcf_2(g_1 + cg_2 + \cdots + c^{L-1}g_L) \equiv 0 \pmod{N}$$
$$2bcf_2(g_1 + bg_2 + \cdots + b^{L-1}g_L) \equiv 0 \pmod{N}$$
$$2c^2 f_2(g_1 + cg_2 + \cdots + c^{L-1}g_L) \equiv 0 \pmod{N}.$$

For the given constituent encoders, we can choose $b = 8$ and $c = 12$, from which it follows that we require $3 \cdot 2^5 f_2 \equiv 0 \pmod{N}$, and the result follows from Proposition 1 and the fact that the Hamming weight of the codeword in Fig. 2 is (at most) 51 when $b = 8$ and $c = 12$. ∎

In Fig. 3, an input-weight 4 codeword is shown. The upper and lower constituent codewords both contain 2 input-weight 2 fundamental paths. The interleaving of the systematic 1-positions is indicated by arrows and the 4 first letters of the English alphabet. To make the pattern in Fig. 3 a codeword, we need to make sure that $f(g(x+a)+b) \equiv f(g(x)+c)+d \pmod{N}$, where $x \in \mathbb{Z}_N$ denotes the first systematic 1-position (labeled by "A" in Fig. 3) of the first fundamental path in the lower constituent codeword. The index differences between the final and first systematic 1-positions of the first and second fundamental paths in the lower constituent codeword are denoted by $a$ and $d$, respectively. The corresponding index differences for the upper constituent codeword are denoted by $c$ and $b$, respectively. With a primitive feedback polynomial of degree $\nu$, $a$, $b$, $c$, and $d$ are all multiples of $2^\nu - 1$. The Hamming weight of the codeword in Fig. 3 is (at most) $12 + 2^{\nu-1}(|a|+|b|+|c|+|d|)/(2^\nu - 1)$. The congruence $f(g(x+a)+b) \equiv f(g(x)+c)+d \pmod{N}$ is equivalent, with $x=0$, to

$$\begin{aligned}(b^2-c^2)\,f_2 + (b-c)\,f_1 + a - d \\ + 2ab f_2\left(g_1 + ag_2 + \cdots + a^{L-1}g_L\right) \equiv 0 \pmod{N}.\end{aligned} \quad (10)$$

If we choose $c=b$ and $d=a$, we get $2abf_2(g_1+ag_2+\cdots+a^{L-1}g_L) \equiv 0 \pmod{N}$. Furthermore, let $a=(2^\nu-1)a'$ and $b=(2^\nu-1)b'$.

In Table I, we list in the second column the congruence in (10) with different values of the pair $(|a'|,|b'|)$ (different rows correspond to different values of the pair $(|a'|,|b'|)$). The third column contains simplified congruences corresponding to the congruences in the second column, in the sense that if a congruence in the third column is satisfied, then also the corresponding congruence in the second column is satisfied. The last column contains an upper bound on the weight of the corresponding codeword in the turbo code.

We will now consider the special case of $\nu=3$ in more detail. Using Proposition 1 and the congruences in Table I, we get the conditions on $N$ and the corresponding upper bounds in Table II. When we assume that the QPP has a quadratic inverse, we can use [20, Theorem 3.6] (or Proposition 3 with $L=2$) to relax the conditions on $N$ in Table II for the same upper bounds on the $d_{\min}$. For instance, the congruence in the fifth row and third column of Table I (with $\nu=3$) is satisfied if $n_{F,2}+3 \geq n_{N,2}$, $n_{F,7}+2 \geq n_{N,7}$, and $n_{F,p} \geq n_{N,p}$, $p \neq 2,7$. Using [20, Theorem 3.6] (or Proposition 3 with $L=2$), it follows that the two inequalities above are satisfied for all values of $f_2$ if

$$n_{N,2} \leq \begin{cases} 3 + \max\left(\left\lceil \frac{n_{N,2}-2}{2}\right\rceil, 1\right), & \text{if } n_{N,2} > 1 \\ 3, & \text{if } n_{N,2}=0,1 \end{cases}$$

$$n_{N,3} \leq \begin{cases} \max\left(\left\lceil \frac{n_{N,3}-1}{2}\right\rceil, 1\right), & \text{if } n_{N,3} > 0 \\ 0, & \text{if } n_{N,3}=0 \end{cases}$$

$$n_{N,7} \leq 2 + \left\lceil \frac{n_{N,7}}{2}\right\rceil$$

$$n_{N,p} \leq \left\lceil \frac{n_{N,p}}{2}\right\rceil, \quad \text{if } p \neq 2,3,7$$

which reduces to

$$n_{N,p} \leq \begin{cases} 5, & \text{if } p=2,7 \\ 1, & \text{otherwise.} \end{cases}$$

In a similar fashion, we can use the congruences in the other rows in Table I together with [20, Theorem 3.6] (or Proposition 3 with $L=2$) to derive other conditions on $N$ for various upper bounds on the optimum minimum distance when the QPP has a quadratic inverse. The conditions and the corresponding upper bounds are summarized in Table III.

### B. General Upper Bound With a Quadratic Inverse

In this section, we provide a general (i.e., independent of $N$) upper bound on the $d_{\min}$ of turbo codes when an inverse polynomial is of degree two, i.e., the inverse permutation is a QPP. In the following, let $g(x) = g_1 x + g_2 x^2 \pmod{N}$ denote the inverse of $f(x)$.

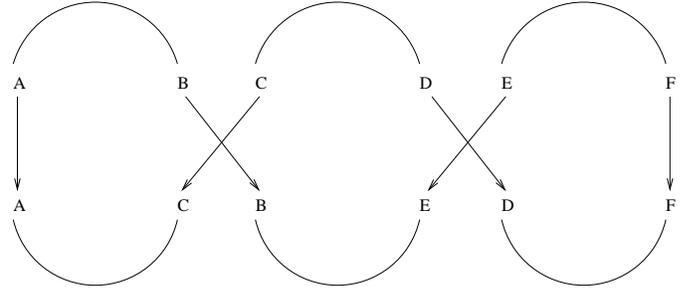

Fig. 4. An input-weight 6 critical codeword for QPP-based interleavers.

*Theorem 3:* The minimum distance of a conventional binary turbo code (of nominal rate $1/3$) using primitive feedback and monic feedforward polynomials of degree $\nu$ and QPPs with a quadratic inverse, is upper-bounded by $2(2^{\nu+1}+9)$.

*Proof:* In Fig. 4, an input-weight 6 codeword is shown. The upper and lower constituent codewords contain 3 input-weight 2 fundamental paths each. The interleaving of the systematic 1-positions is indicated by arrows and the 6 first letters of the English alphabet. To make the pattern in Fig. 4 a codeword, we need to make sure that $f(g(f(x)+a)+2a) \equiv f(g(f(x+a)+2a)+a) - a \pmod{N}$, where $x \in \mathbb{Z}_N$ is the leftmost systematic 1-position (labeled by "A" in Fig. 4) in the upper constituent codeword. The index difference between the final and first systematic 1-positions of the first and third fundamental paths in both the upper and lower constituent codewords is denoted by $a$. The corresponding index difference for the second (i.e., middle) fundamental path (in both constituent codewords) is $2a$. Since the feedback polynomial is primitive, we can choose $a = 2^\nu - 1$, and with a monic feedforward polynomial of degree $\nu$, the parity weight of the first and third fundamental paths in both constituent codewords in Fig. 4 is $2+2^{\nu-1}$. The corresponding parity weight for the second (i.e., middle) fundamental path (in both constituent codewords) is $2+2^\nu$. Now, the congruence above is equivalent to

$$4 \cdot a^3 \cdot f_2 g_2(1 + 2f_1 + 2af_2 + 4xf_2) \equiv 0 \pmod{N}$$

which again is equivalent to

$$4 \cdot a^3 \cdot f_2 g_2 (1 + 2f_1 + 2af_2) \equiv 0 \pmod{N} \quad (11)$$

since $4f_2^2 g_2 \equiv 0 \pmod{N}$. This follows from [20, Theorem 3.5] which states that $12 f_2 g_2 \equiv 0 \pmod{N}$. The Hamming weight of the codeword in Fig. 4 is (at most) $2(2^{\nu+1}+9)$, where equality holds if the fundamental paths do not interfere with each other as in Fig. 4.

From [20, Theorem 3.5] we know that $12 f_2 g_2 \equiv 0 \pmod{N}$, and it follows that $4 f_2 g_2 \equiv 0 \pmod{N}$ if 27 is not a divisor of $N$. Thus, the congruence in (11) holds for *all* QPPs with a quadratic inverse, for a given value of $N$, if 27 is *not* a divisor of $N$.

In Fig. 5, another input-weight 6 codeword is depicted. The upper and lower constituent codewords contain 3 input-weight

TABLE I
SUMMARY OF CONDITIONS THAT MAKE THE PATTERN IN FIG. 3 A CODEWORD.

| $(|a'|,|b'|)$ | Equation (10) with $c=b$ and $d=a$ | Simplified congruence | Weight |
|---|---|---|---|
| $(1,1)$ | $(2^\nu-1)^2 \cdot 2f_2(g_1 \pm (2^\nu-1)g_2 + \cdots + (\pm(2^\nu-1))^{L-1}g_L) \equiv 0$ | $(2^\nu-1)^2 \cdot 2f_2 \equiv 0$ | $12+2^{\nu+1}$ |
| $(1,2)$ | $(2^\nu-1)^2 \cdot 2^2 f_2(g_1 \pm (2^\nu-1)g_2 + \cdots + (\pm(2^\nu-1))^{L-1}g_L) \equiv 0$ | $(2^\nu-1)^2 \cdot 2^2 f_2 \equiv 0$ | $12+3\cdot 2^\nu$ |
| $(2,1)$ | $(2^\nu-1)^2 \cdot 2^2 f_2(g_1 \pm 2(2^\nu-1)g_2 + \cdots + (\pm 2(2^\nu-1))^{L-1}g_L) \equiv 0$ | $(2^\nu-1)^2 \cdot 2^2 f_2 \equiv 0$ | $12+3\cdot 2^\nu$ |
| $(2,2)$ | $(2^\nu-1)^2 \cdot 2^3 f_2(g_1 \pm 2(2^\nu-1)g_2 + \cdots + (\pm 2(2^\nu-1))^{L-1}g_L) \equiv 0$ | $(2^\nu-1)^2 \cdot 2^3 f_2 \equiv 0$ | $12+2^{\nu+2}$ |
| $(1,3)$ | $(2^\nu-1)^2 \cdot 2\cdot 3 f_2(g_1 \pm (2^\nu-1)g_2 + \cdots + (\pm(2^\nu-1))^{L-1}g_L) \equiv 0$ | $(2^\nu-1)^2 \cdot 2\cdot 3 f_2 \equiv 0$ | $12+2^{\nu+2}$ |
| $(3,1)$ | $(2^\nu-1)^2 \cdot 2\cdot 3 f_2(g_1 \pm 3(2^\nu-1)g_2 + \cdots + (\pm 3(2^\nu-1))^{L-1}g_L) \equiv 0$ | $(2^\nu-1)^2 \cdot 2\cdot 3 f_2 \equiv 0$ | $12+2^{\nu+2}$ |

TABLE II
SUMMARY OF CONDITIONS ON $N$ (DERIVED USING THE CONGRUENCES IN TABLE I) FOR VARIOUS UPPER BOUNDS ON THE OPTIMUM MINIMUM DISTANCE WHEN $\nu=3$.

| $n_{N,2}$ | $n_{N,3}$ | $n_{N,7}$ | $n_{N,p}, p\neq 2,3,7$ | Upper bound | Row from Table I |
|---|---|---|---|---|---|
| $\leq 2$ | $\leq 1$ | $\leq 3$ | $\leq 1$ | 28 | 2 |
| $\leq 3$ | $\leq 1$ | $\leq 3$ | $\leq 1$ | 36 | 3 and 4 |
| $\leq 4$ | $\leq 1$ | $\leq 3$ | $\leq 1$ | 44 | 5 |
| $\leq 2$ | $\leq 2$ | $\leq 3$ | $\leq 1$ | 44 | 6 and 7 |

TABLE III
SUMMARY OF CONDITIONS ON $N$ (DERIVED USING THE CONGRUENCES IN TABLE I) FOR VARIOUS UPPER BOUNDS ON THE OPTIMUM MINIMUM DISTANCE WHEN $\nu=3$ AND THE INVERSE DEGREE IS TWO.

| $n_{N,2}$ | $n_{N,3}$ | $n_{N,7}$ | $n_{N,p}, p\neq 2,3,7$ | Upper bound | Row from Table I |
|---|---|---|---|---|---|
| $\leq 2$ | $\leq 1$ | $\leq 5$ | $\leq 1$ | 28 | 2 |
| $\leq 3$ | $\leq 1$ | $\leq 5$ | $\leq 1$ | 36 | 3 and 4 |
| $\leq 5$ | $\leq 1$ | $\leq 5$ | $\leq 1$ | 44 | 5 |
| $\leq 2$ | $\leq 2$ | $\leq 5$ | $\leq 1$ | 44 | 6 and 7 |

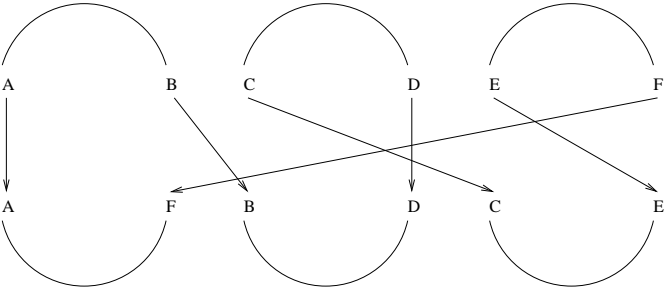

Fig. 5. An input-weight 6 critical codeword for QPP-based interleavers.

2 fundamental paths each. The interleaving of the systematic 1-positions is indicated by arrows and the 6 first letters of the English alphabet. To make this pattern a codeword, $f(g(f(x) + 2a) - a) \equiv f(g(f(x+a) + a) - 2a) + a \pmod{N}$, where $x \in \mathbb{Z}_N$ is the leftmost systematic 1-position (labeled by "A" in Fig. 5) in the upper constituent codeword. The index difference between the final and first systematic 1-positions of the first and third fundamental paths in the upper constituent codeword, and of the second and third fundamental paths in the lower constituent codeword, is denoted by $a$. The corresponding index difference for the second fundamental path in the upper constituent codeword, and for the first fundamental path in the lower constituent codeword is $2a$. Now, the congruence above is equivalent to

$$4 \cdot a^3 \cdot f_2 g_2(1 - 2f_1 - 2af_2 - 4xf_2) \equiv 0 \pmod{N}$$

which again is equivalent to

$$4 \cdot a^3 \cdot f_2 g_2(1 - 2f_1 - 2af_2) \equiv 0 \pmod{N} \qquad (12)$$

since $4f_2^2 g_2 \equiv 0 \pmod{N}$. This follows from [20, Theorem 3.5] which states that $12 f_2 g_2 \equiv 0 \pmod{N}$. As for the codeword in Fig. 4, the Hamming weight of the codeword in Fig. 5 is (at most) $2(2^{\nu+1}+9)$. Assume $27|N$. Then, $f_2 = 3 \cdot c$, for some integer $c$, since $3|f_2$. Furthermore, $f_1 = 1 + 3 \cdot k$ or $2 + 3 \cdot k$, for some integer $k$, since $\gcd(f_1 + f_2, N) = 1$. If $f_1 = 1 + 3 \cdot k$, then the congruence in (11) reduces to $4 \cdot a^3 \cdot f_2 g_2(1 + 2(1+3\cdot k) + 2a \cdot 3 \cdot c) = 12 \cdot a^3 \cdot f_2 g_2(1 + 2\cdot k + 2a \cdot c) \equiv 0 \pmod{N}$, which is always true, since $12 f_2 g_2 \equiv 0 \pmod{N}$ [20, Theorem 3.5]. Furthermore, if $f_1 = 2 + 3 \cdot k$, then the congruence in (12) reduces to $4 \cdot a^3 \cdot f_2 g_2(1 - 2(2+3\cdot k) - 2a \cdot 3 \cdot c) = 12 \cdot a^3 \cdot f_2 g_2(-1 - 2\cdot k - 2a \cdot c) \equiv 0 \pmod{N}$, which is always true, since $12 f_2 g_2 \equiv 0 \pmod{N}$ [20, Theorem 3.5]. Thus, there is an upper bound of $2(2^{\nu+1}+9)$ on the $d_{\min}$ for QPPs with a quadratic inverse for all values of $N$. ∎

We emphasis that Theorem 3 applies for all interleaver lengths $N$ and is achievable, at least for $\nu=3$, for a range of $N$-values (see Fig 8). Thus, for $\nu = 3$, to achieve a $d_{\min}$ strictly larger than $2(2^{\nu+1} + 9) = 50$, the degree of an inverse permutation should be at least three.

*Example 2:* For $N = 5504 = 2^7 \times 43$, it follows from Theorem 3 (with $\nu = 3$) that the $d_{\min}$ is at most 50 when the QPP has a quadratic inverse. Furthermore, using the condition in (1) in Example 1 for the QPP to have a cubic inverse, we observe that the condition $3 \cdot 2^5 f_2 \equiv 0 \pmod{N}$ from the proof of Theorem 2 is always satisfied, which means that to achieve a $d_{\min}$ (with the 3GPP LTE encoders) strictly larger than 51, the degree of an inverse polynomial has to be at least four.

### C. Partial Upper Bounds With a Quadratic Inverse

*Theorem 4:* The minimum distance of a conventional binary turbo code (of nominal rate $1/3$) with feedback polynomial $1 + D^2 + D^3$ and feedforward polynomial $1 + D + D^3$, is upper-bounded by $38 + 12l$, where $l$ is some nonnegative integer, for all interleaver lengths $N$ that satisfy

$$n_{N,p} \leq \begin{cases} 2l + 5, & \text{if } p = 2 \\ 3, & \text{if } p = 7 \\ 1, & \text{otherwise} \end{cases} \quad (13)$$

when using QPP interleavers with a quadratic inverse.

*Proof:* The proof is based on the codeword in Fig 1. In the case that an inverse polynomial is of degree two, the congruences in (4) and (5) reduce (with $L = 2$, $a = 2^l \cdot 7$, $b = 8$, and $c = 12$) to $2^l \cdot 112 f_2(g_1 + 8g_2) \equiv 0 \pmod{N}$ and $2^l \cdot 168 f_2(g_1 + 12g_2) \equiv 2^l \cdot 168 f_2 g_1 \equiv 0 \pmod{N}$ [20, Theorem 3.5]. Furthermore, if $2^l \cdot 168 f_2 g_1 \equiv 0 \pmod{N}$, then $2^l(2 \cdot 168 f_2 g_1 + 224 \cdot 12 f_2 g_2) = 2^l(3 \cdot 112 f_2(g_1 + 8g_2)) \equiv 0 \pmod{N}$ [20, Theorem 3.5], from which it follows that $2^l \cdot 112 f_2(g_1 + 8g_2) \equiv 0 \pmod{N}$ if 9 is not a divisor of $N$. The congruence $2^l \cdot 168 f_2 g_1 \equiv 0 \pmod{N}$ is satisfied (for all valid values of $g_1$ and $f_2$) if $n_{F,2} + l + 3 \geq n_{N,2}$, $n_{F,3} + 1 \geq n_{N,3}$, $n_{F,7} + 1 \geq n_{N,7}$, and $n_{F,p} \geq n_{N,p}$, $p \neq 2, 3, 7$. Using [20, Theorem 3.6] (or Proposition 3 with $L = 2$), it follows that the congruences are satisfied for all valid values of $f_2$, for a given value of $N$, if the following conditions are satisfied.

$$n_{N,2} \leq \begin{cases} l + 3 + \max\left(\left\lceil \frac{n_{N,2}-2}{2} \right\rceil, 1\right), & \text{if } n_{N,2} > 1 \\ l + 3, & \text{if } n_{N,2} = 0, 1 \end{cases}$$

$$n_{N,3} \leq \begin{cases} 1 + \max\left(\left\lceil \frac{n_{N,3}-1}{2} \right\rceil, 1\right), & \text{if } n_{N,3} > 0 \\ 1, & \text{if } n_{N,3} = 0 \end{cases}$$

$$n_{N,7} \leq 1 + \left\lceil \frac{n_{N,7}}{2} \right\rceil$$

$$n_{N,p} \leq \left\lceil \frac{n_{N,p}}{2} \right\rceil, \quad \text{if } p \neq 2, 3, 7$$

which reduce (when we use the fact that 9 should not be a divisor of $N$) to (13). The Hamming weight of the codeword in Fig. 1 is (at most) $38 + 12l$, and the result follows. ∎

We remark that Theorem 4 is only useful when $l = 0$, since the $d_{\min}$ upper bound is at least 50 when $l \geq 1$. This is the case since Theorem 3 (with $\nu = 3$) gives a general upper bound of 50.

### D. Partial Upper Bounds With an Inverse Degree of Three

*Theorem 5:* The minimum distance of a conventional binary turbo code (of nominal rate $1/3$) with feedback polynomial $1 + D^2 + D^3$ and feedforward polynomial $1 + D + D^3$, is upper-bounded by $38 + 12l$, where $l$ is some nonnegative integer, for all interleaver lengths $N$ that satisfy

$$n_{N,p} \leq \begin{cases} \left\lfloor \frac{3l}{2} \right\rfloor + 4, & \text{if } p = 2 \\ 2, & \text{if } p = 7 \\ 1, & \text{otherwise} \end{cases} \quad (14)$$

when using QPP interleavers with a cubic inverse.

*Proof:* The result can be proved using the same arguments as in the proof of Theorem 4, but using Proposition 3 with $L = 3$ instead of [20, Theorem 3.6] (or Proposition 3 with $L = 2$). In particular, we require that

$$n_{N,2} \leq \begin{cases} l + 3 + \max\left(\left\lceil \frac{n_{N,2}-3}{3} \right\rceil, 1\right), & \text{if } n_{N,2} > 1 \\ l + 3, & \text{if } n_{N,2} \leq 1 \end{cases}$$

$$n_{N,3} \leq \begin{cases} 1 + \max\left(\left\lceil \frac{n_{N,3}-1}{3} \right\rceil, 1\right), & \text{if } n_{N,3} > 0 \\ 1, & \text{if } n_{N,3} = 0 \end{cases}$$

$$n_{N,5} \leq \begin{cases} \max\left(\left\lceil \frac{n_{N,5}-1}{3} \right\rceil, 1\right), & \text{if } n_{N,5} > 0 \\ 0, & \text{if } n_{N,5} = 0 \end{cases}$$

$$n_{N,7} \leq 1 + \left\lceil \frac{n_{N,7}}{3} \right\rceil$$

$$n_{N,p} \leq \left\lceil \frac{n_{N,p}}{3} \right\rceil, \quad \text{if } p \neq 2, 3, 5, 7$$

which gives the upper bound in (14). ∎

*Theorem 6:* The minimum distance of a conventional binary turbo code (of nominal rate $1/3$) using primitive feedback and monic feedforward polynomials of degree $\nu$ and QPPs with an inverse degree of three, is upper-bounded by $2(2^{\nu+1} + 9)$ for all interleaver lengths $N$ that satisfy

$$n_{N,p} \leq \begin{cases} 4, & \text{if } p = 2 \\ \left\lfloor \frac{9 n_{2^\nu - 1, p}}{2} \right\rfloor + 2, & \text{if } p = 3, 5 \\ \left\lceil \frac{9 n_{2^\nu - 1, p}}{2} \right\rceil + 2, & \text{otherwise.} \end{cases} \quad (15)$$

*Proof:* By repeating the first part of the proof of Theorem 3 with $g(x) = g_1 x + g_2 x^2 + g_3 x^3$, the congruence in (11) (with $x = 0$) is generalized to

$$4a^3 f_2 \left(g_2 (1 + 2f_1 + 2a f_2) \right. \\ \left. + 3 g_3 a (1 + f_1 + a f_2)^2 \right) \equiv 0 \pmod{N}. \quad (16)$$

If 4 is a factor in $N$, then 2 is a factor in both $g_2$ and $g_3$, and if 3 is a factor in $N$, then 3 is also a factor in $g_2$ [12]. Furthermore, if $p > 3$ is at least a double factor in $N$, then $p$ is a factor in both $g_2$ and $g_3$ [12]. Combining this with the congruence in (16), we require that (see Proposition 3 with

$L = 3$ for details)

$$n_{N,2} \leq \begin{cases} 3 + \max\left(\left\lceil \frac{n_{N,2}-3}{3} \right\rceil, 1\right), & \text{if } n_{N,2} > 1 \\ 3, & \text{if } n_{N,2} \leq 1 \end{cases}$$

$$n_{N,3} \leq \begin{cases} 1 + 3n_{2^\nu-1,3} + \max\left(\left\lceil \frac{n_{N,3}-1}{3} \right\rceil, 1\right), & \text{if } n_{N,3} > 0 \\ 1 + 3n_{2^\nu-1,3}, & \text{if } n_{N,3} = 0 \end{cases}$$

$$n_{N,5} \leq \begin{cases} 1 + 3n_{2^\nu-1,5} + \max\left(\left\lceil \frac{n_{N,5}-1}{3} \right\rceil, 1\right), & \text{if } n_{N,5} > 0 \\ 1 + 3n_{2^\nu-1,5}, & \text{if } n_{N,5} = 0 \end{cases}$$

$$n_{N,p} \leq 1 + 3n_{2^\nu-1,p} + \left\lceil \frac{n_{N,p}}{3} \right\rceil, \quad \text{if } p \neq 2, 3, 5$$

which simplifies to the result in (15). ∎

*E. Upper Bounds With Dual Termination*

We remark that the upper bounds on the $d_{\min}$ derived above can be shown to hold with dual termination as well as long as $N$ is sufficient large. As an example, we consider the upper bound of Theorem 3, which holds with dual termination when $N \geq 2^{\nu+3} - 7$. The argument is as follows.

The bound in Theorem 3 is based on the upper and lower constituent codewords in Figs. 4 and 5. For details, see the proof of Theorem 3. Now, consider the codeword in Fig. 4. Let $x \in \mathbb{Z}_N$ denote the leftmost systematic 1-position in the upper constituent codeword. For a given value of $x$, the fundamental paths in Fig. 4 may wrap around at the end of the trellis. Since all the systematic 1-positions are determined by the value of $x$, there will be exactly $L_i - 1$ values for $x$ that will make the $i$th, $i = 0, \ldots, Q-1$, fundamental path wrap around at the end of the trellis, where $L_i$ is the length of the $i$th fundamental path and $Q$ is the total number of fundamental paths in the upper and lower constituent codewords. By repeating the argument, we get that there will be at most $L - Q$ values for $x$ that will make at least one of the fundamental paths wrap around at the end of the trellis, where $L = \sum_{i=0}^{Q-1} L_i$, and we get the condition $N - (L - Q) \geq 1$, which simplifies to $N \geq 2^{\nu+3} - 7$, since, for the codeword in Fig. 4, $Q = 6$ and $L = 8 \cdot 2^\nu - 2$.

Note that the argument above can be repeated for the codeword in Fig. 5, and we get exactly the same lower bound on $N$. Thus, if $N \geq 2^{\nu+3} - 7$, there will exist at least one value for $x$ such that none of the fundamental paths in Figs. 4 and 5 will wrap around at the end of the trellis, and the result follows.

IV. 3GPP LTE TURBO CODES

The turbo codes in the 3GPP LTE standard [15] use the 8-state constituent encoder with feedforward polynomial $1 + D + D^3$ and feedback polynomial $1 + D^2 + D^3$. We have computed the *exact* $d_{\min}$ and the corresponding multiplicity of the 3GPP LTE turbo codes (with dual termination) for all interleaver lengths $N$ using the algorithm from [22]. To speed-up the computation, we have used the fact that the 3GPP LTE turbo codes are quasi-cyclic (with tailbiting termination) with period $N/\gcd(2f_2, N)$. The results with dual termination are presented in Table IV. With dual termination both constituent trellises are terminated to the zero state and all systematic bits are included in the interleaver. Due to trellis termination some of the systematic bits (twice the constraint length) are redundant and these bits are in general not consecutive at the end of the input block. We have also tailored a version of the triple impulse method [9] which also explicitly checks for special low input-weight codewords. We ran the method using all the 3GPP LTE QPPs of length at least 512. For all lengths, this method found the *exact* $d_{\min}$ and also the *exact* multiplicity in all but the case of $N = 1920$, where the multiplicity was slightly underestimated. Finally, we remark that the 3GPP LTE specification [15] uses a different trellis termination technique, and thus the results in Table IV do not strictly apply to the codes from the standard.

For the 3GPP LTE lengths

$$\{496, 624, 656, 688, 752, 816, 848, 880, 912, 944, 976\}$$

the results in Table IV together with the upper bound from Theorem 1 (with $l = 0$) show that the 3GPP LTE interleavers are $d_{\min}$-optimal, i.e., there are no QPPs that give a *strictly* larger $d_{\min}$, for these lengths.

In a similar fashion, for the 3GPP LTE lengths

$$\{1696, 1760, 1952\}$$

the results in Table IV together with the upper bound from Theorem 1 (with $l = 1$) show that the 3GPP LTE interleavers are $d_{\min}$-optimal for these lengths.

The 3GPP LTE lengths greater than 2048 that satisfy the conditions of Theorem 2 are

$$\{2112, 2240, 2368, 2496, 2624, 2752, 2880, 3008, 3264, 3392,$$
$$3520, 3648, 3776, 3904, 4032, 4160, 4288, 4416, 4544, 4672,$$
$$4928, 5056, 5312, 5440, 5568, 5696, 5824, 5952, 6080\}.$$

From the results of Table IV, it follows that the 3GPP LTE interleavers are $d_{\min}$-optimal for lengths

$$\{4288, 4544, 5056, 5312, 5568, 6080\}.$$

Finally, for the lengths

$$\{1920, 2176, 2368, 2432, 2496, 2624, 2752, 3264, 3392, 3456,$$
$$3520, 3648, 3712, 3776, 3840, 3904, 3968, 4224, 4864, 5376,$$
$$5632, 5760, 5888, 6144\}$$

the 3GPP LTE QPPs have a quadratic inverse, and thus (from Theorem 3 with $\nu = 3$) they are $d_{\min}$-optimal within the class of QPPs with a quadratic inverse, since they give a $d_{\min}$ of 50 (see Table IV).

V. COMPUTER SEARCH

In Fig. 6, the optimum $d_{\min}$ within the class of all irreducible QPPs (found by computer search) for 8-state turbo codes is plotted versus the interleaver length $N$. In particular, we have considered the 3GPP LTE interleaver lengths from 40 to 1008, i.e., the interleaver lengths $40, 48, \ldots, 512, 528, 544, \ldots, 1008$. Note that in some cases, a better $d_{\min}$ may be achieved by using an LPP, i.e., a reducible QPP, for small interleaver lengths [16], but the best achievable minimum distance is upper-bounded by 27 over the class of reducible QPPs [16]. For comparison, we also show

TABLE IV
THE EXACT MINIMUM DISTANCE $d_{\min}$ AND ITS CORRESPONDING MULTIPLICITY $N_{d_{\min}}$ OF 3GPP LTE TURBO CODES (WITH DUAL TERMINATION).

| $N$ | $f(x)$ | $d_{\min}$ | $N_{d_{\min}}$ | $N$ | $f(x)$ | $d_{\min}$ | $N_{d_{\min}}$ | $N$ | $f(x)$ | $d_{\min}$ | $N_{d_{\min}}$ |
|---|---|---|---|---|---|---|---|---|---|---|---|
| 40 | $3x + 10x^2$ | 17 | 11 | 576 | $65x + 96x^2$ | 38 | 1078 | 2240 | $209x + 420x^2$ | 44 | 1092 |
| 48 | $7x + 12x^2$ | 17 | 16 | 592 | $19x + 74x^2$ | 37 | 136 | 2304 | $253x + 216x^2$ | 44 | 561 |
| 56 | $19x + 42x^2$ | 14 | 23 | 608 | $37x + 76x^2$ | 38 | 1130 | 2368 | $367x + 444x^2$ | 50 | 9235 |
| 64 | $7x + 16x^2$ | 20 | 22 | 624 | $41x + 234x^2$ | 38 | 1162 | 2432 | $265x + 456x^2$ | 50 | 9491 |
| 72 | $7x + 18x^2$ | 23 | 51 | 640 | $39x + 80x^2$ | 38 | 1194 | 2496 | $181x + 468x^2$ | 50 | 9748 |
| 80 | $11x + 20x^2$ | 23 | 103 | 656 | $185x + 82x^2$ | 38 | 1226 | 2560 | $39x + 80x^2$ | 44 | 627 |
| 88 | $5x + 22x^2$ | 23 | 32 | 672 | $43x + 252x^2$ | 30 | 157 | 2624 | $27x + 164x^2$ | 50 | 10259 |
| 96 | $11x + 24x^2$ | 21 | 36 | 688 | $21x + 86x^2$ | 38 | 1290 | 2688 | $127x + 504x^2$ | 38 | 658 |
| 104 | $7x + 26x^2$ | 27 | 114 | 704 | $155x + 44x^2$ | 39 | 330 | 2752 | $143x + 172x^2$ | 50 | 10771 |
| 112 | $41x + 84x^2$ | 22 | 171 | 720 | $79x + 120x^2$ | 38 | 1366 | 2816 | $43x + 88x^2$ | 44 | 690 |
| 120 | $103x + 90x^2$ | 26 | 44 | 736 | $139x + 92x^2$ | 38 | 1386 | 2880 | $29x + 300x^2$ | 39 | 118 |
| 128 | $15x + 32x^2$ | 21 | 51 | 752 | $23x + 94x^2$ | 38 | 1418 | 2944 | $45x + 92x^2$ | 44 | 722 |
| 136 | $9x + 34x^2$ | 28 | 103 | 768 | $217x + 48x^2$ | 39 | 180 | 3008 | $157x + 188x^2$ | 47 | 369 |
| 144 | $17x + 108x^2$ | 30 | 205 | 784 | $25x + 98x^2$ | 30 | 186 | 3072 | $47x + 96x^2$ | 44 | 753 |
| 152 | $9x + 38x^2$ | 28 | 60 | 800 | $17x + 80x^2$ | 45 | 1652 | 3136 | $13x + 28x^2$ | 44 | 1540 |
| 160 | $21x + 120x^2$ | 31 | 248 | 816 | $127x + 102x^2$ | 38 | 1546 | 3200 | $111x + 240x^2$ | 49 | 157 |
| 168 | $101x + 84x^2$ | 27 | 592 | 832 | $25x + 52x^2$ | 39 | 197 | 3264 | $443x + 204x^2$ | 50 | 12819 |
| 176 | $21x + 44x^2$ | 28 | 71 | 848 | $239x + 106x^2$ | 38 | 1610 | 3328 | $51x + 104x^2$ | 44 | 818 |
| 184 | $57x + 46x^2$ | 29 | 144 | 864 | $17x + 48x^2$ | 36 | 824 | 3392 | $51x + 212x^2$ | 50 | 13331 |
| 192 | $23x + 48x^2$ | 31 | 148 | 880 | $137x + 110x^2$ | 38 | 1674 | 3456 | $451x + 192x^2$ | 50 | 6802 |
| 200 | $13x + 50x^2$ | 31 | 81 | 896 | $215x + 112x^2$ | 38 | 2126 | 3520 | $257x + 220x^2$ | 50 | 13843 |
| 208 | $27x + 52x^2$ | 31 | 181 | 912 | $29x + 114x^2$ | 38 | 1738 | 3584 | $57x + 336x^2$ | 38 | 440 |
| 216 | $11x + 36x^2$ | 29 | 61 | 928 | $15x + 58x^2$ | 42 | 110 | 3648 | $313x + 228x^2$ | 50 | 14355 |
| 224 | $27x + 56x^2$ | 22 | 99 | 944 | $147x + 118x^2$ | 38 | 1802 | 3712 | $271x + 232x^2$ | 50 | 14611 |
| 232 | $85x + 58x^2$ | 33 | 92 | 960 | $29x + 60x^2$ | 39 | 227 | 3776 | $179x + 236x^2$ | 50 | 14867 |
| 240 | $29x + 60x^2$ | 25 | 101 | 976 | $59x + 122x^2$ | 38 | 1866 | 3840 | $331x + 120x^2$ | 50 | 7571 |
| 248 | $33x + 62x^2$ | 35 | 212 | 992 | $65x + 124x^2$ | 38 | 1898 | 3904 | $363x + 244x^2$ | 50 | 15379 |
| 256 | $15x + 32x^2$ | 30 | 53 | 1008 | $55x + 84x^2$ | 30 | 322 | 3968 | $375x + 248x^2$ | 50 | 15635 |
| 264 | $17x + 198x^2$ | 35 | 339 | 1024 | $31x + 64x^2$ | 43 | 121 | 4032 | $127x + 168x^2$ | 38 | 663 |
| 272 | $33x + 68x^2$ | 36 | 1633 | 1056 | $17x + 66x^2$ | 42 | 125 | 4096 | $31x + 64x^2$ | 44 | 504 |
| 280 | $103x + 210x^2$ | 22 | 126 | 1088 | $171x + 204x^2$ | 44 | 516 | 4160 | $33x + 130x^2$ | 49 | 1026 |
| 288 | $19x + 36x^2$ | 36 | 122 | 1120 | $67x + 140x^2$ | 38 | 2953 | 4224 | $43x + 264x^2$ | 50 | 16659 |
| 296 | $19x + 74x^2$ | 35 | 260 | 1152 | $35x + 72x^2$ | 43 | 138 | 4288 | $33x + 134x^2$ | 51 | 4484 |
| 304 | $37x + 76x^2$ | 36 | 1606 | 1184 | $19x + 74x^2$ | 43 | 141 | 4352 | $477x + 408x^2$ | 44 | 1073 |
| 312 | $19x + 78x^2$ | 36 | 1654 | 1216 | $39x + 76x^2$ | 44 | 580 | 4416 | $35x + 138x^2$ | 49 | 1090 |
| 320 | $21x + 120x^2$ | 32 | 68 | 1248 | $19x + 78x^2$ | 43 | 299 | 4480 | $233x + 280x^2$ | 46 | 550 |
| 328 | $21x + 82x^2$ | 35 | 292 | 1280 | $199x + 240x^2$ | 44 | 612 | 4544 | $357x + 142x^2$ | 51 | 4476 |
| 336 | $115x + 84x^2$ | 30 | 147 | 1312 | $21x + 82x^2$ | 45 | 632 | 4608 | $337x + 480x^2$ | 44 | 758 |
| 344 | $193x + 86x^2$ | 36 | 1846 | 1344 | $211x + 252x^2$ | 30 | 163 | 4672 | $37x + 146x^2$ | 49 | 1155 |
| 352 | $21x + 44x^2$ | 35 | 77 | 1376 | $21x + 86x^2$ | 47 | 660 | 4736 | $71x + 444x^2$ | 51 | 4668 |
| 360 | $133x + 90x^2$ | 33 | 159 | 1408 | $43x + 88x^2$ | 44 | 676 | 4800 | $71x + 120x^2$ | 49 | 472 |
| 368 | $81x + 46x^2$ | 35 | 82 | 1440 | $149x + 60x^2$ | 44 | 461 | 4864 | $37x + 152x^2$ | 50 | 9619 |
| 376 | $45x + 94x^2$ | 35 | 168 | 1472 | $45x + 92x^2$ | 44 | 708 | 4928 | $39x + 462x^2$ | 38 | 304 |
| 384 | $23x + 48x^2$ | 33 | 85 | 1504 | $49x + 846x^2$ | 47 | 180 | 4992 | $127x + 234x^2$ | 47 | 155 |
| 392 | $243x + 98x^2$ | 30 | 175 | 1536 | $71x + 48x^2$ | 35 | 93 | 5056 | $39x + 158x^2$ | 51 | 5300 |
| 400 | $151x + 40x^2$ | 39 | 499 | 1568 | $13x + 28x^2$ | 38 | 3050 | 5120 | $39x + 80x^2$ | 44 | 632 |
| 408 | $155x + 102x^2$ | 35 | 372 | 1600 | $17x + 80x^2$ | 45 | 1709 | 5184 | $31x + 96x^2$ | 44 | 6839 |
| 416 | $25x + 52x^2$ | 38 | 838 | 1632 | $25x + 102x^2$ | 35 | 200 | 5248 | $113x + 902x^2$ | 45 | 162 |
| 424 | $51x + 106x^2$ | 33 | 187 | 1664 | $183x + 104x^2$ | 49 | 3621 | 5312 | $41x + 166x^2$ | 51 | 5244 |
| 432 | $47x + 72x^2$ | 35 | 129 | 1696 | $55x + 954x^2$ | 50 | 3264 | 5376 | $251x + 336x^2$ | 50 | 21267 |
| 440 | $91x + 110x^2$ | 36 | 2422 | 1728 | $127x + 96x^2$ | 43 | 374 | 5440 | $43x + 170x^2$ | 49 | 1346 |
| 448 | $29x + 168x^2$ | 22 | 105 | 1760 | $27x + 110x^2$ | 50 | 3392 | 5504 | $21x + 86x^2$ | 51 | 340 |
| 456 | $29x + 114x^2$ | 35 | 420 | 1792 | $29x + 112x^2$ | 30 | 219 | 5568 | $43x + 174x^2$ | 51 | 5500 |
| 464 | $247x + 58x^2$ | 34 | 106 | 1824 | $29x + 114x^2$ | 45 | 888 | 5632 | $45x + 176x^2$ | 50 | 11155 |
| 472 | $29x + 118x^2$ | 33 | 213 | 1856 | $57x + 116x^2$ | 44 | 900 | 5696 | $45x + 178x^2$ | 49 | 1411 |
| 480 | $89x + 180x^2$ | 38 | 874 | 1888 | $45x + 354x^2$ | 47 | 686 | 5760 | $161x + 120x^2$ | 50 | 11411 |
| 488 | $91x + 122x^2$ | 36 | 2710 | 1920 | $31x + 120x^2$ | 50 | 7443 | 5824 | $89x + 182x^2$ | 38 | 360 |
| 496 | $157x + 62x^2$ | 38 | 906 | 1952 | $59x + 610x^2$ | 50 | 3776 | 5888 | $323x + 184x^2$ | 50 | 11667 |
| 504 | $55x + 84x^2$ | 30 | 308 | 1984 | $185x + 124x^2$ | 44 | 964 | 5952 | $47x + 186x^2$ | 49 | 1476 |
| 512 | $31x + 64x^2$ | 33 | 117 | 2016 | $113x + 420x^2$ | 30 | 329 | 6016 | $23x + 94x^2$ | 49 | 185 |
| 528 | $17x + 66x^2$ | 37 | 120 | 2048 | $31x + 64x^2$ | 44 | 497 | 6080 | $47x + 190x^2$ | 51 | 6012 |
| 544 | $35x + 68x^2$ | 38 | 1002 | 2112 | $17x + 66x^2$ | 43 | 130 | 6144 | $263x + 480x^2$ | 50 | 12179 |
| 560 | $227x + 420x^2$ | 36 | 3142 | 2176 | $171x + 136x^2$ | 50 | 8467 | | | | |

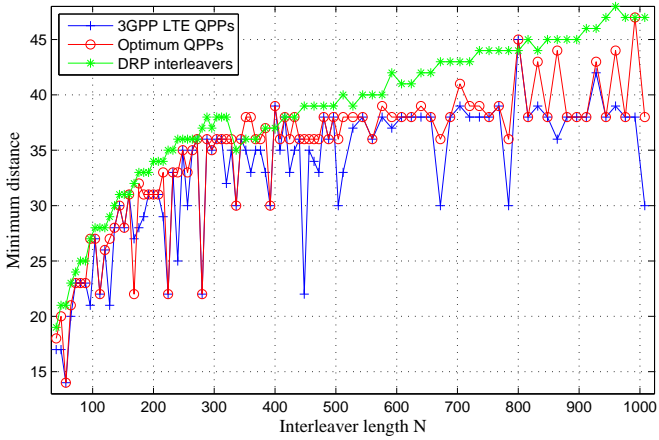

Fig. 6. Comparison of the minimum distance with the 3GPP LTE QPPs with the minimum distance with optimum QPPs, and with DRP interleavers for the 3GPP LTE lengths from 40 to 1008 for 8-state turbo codes.

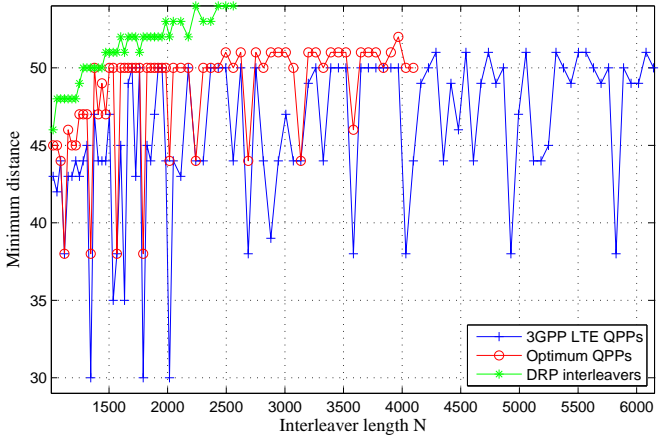

Fig. 7. Comparison of the minimum distance with the 3GPP LTE QPPs with the minimum distance with optimum QPPs, and with DRP interleavers for the 3GPP LTE lengths from 1024 to 4096 for 8-state turbo codes.

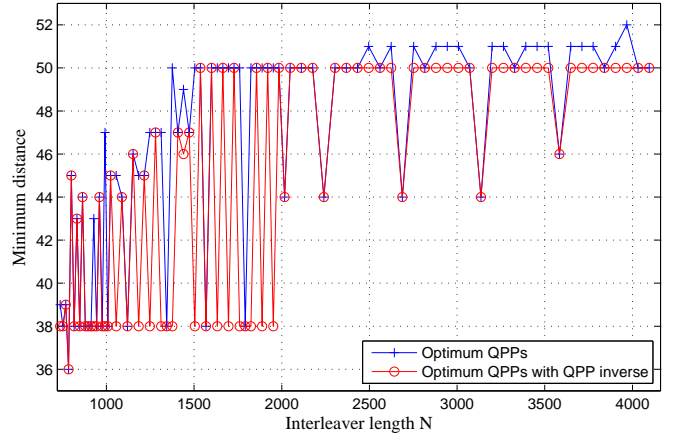

Fig. 8. Comparison of the minimum distance with general optimum QPPs with the minimum distance with optimum QPPs with a quadratic inverse for the 3GPP LTE lengths from 736 to 4096 for 8-state turbo codes.

the results with optimized DRP interleavers [9, 23] and the results with the 3GPP LTE QPPs from Table IV. Note that the $d_{\min}$ values with the DRP interleavers are only estimates for $N > 800$ [9, 23]. In the search, to quickly reject bad QPPs, we used the special version of the triple impulse method mentioned above. In the final stage, the best candidate QPPs were checked using the exhaustive algorithm from [22]. In Fig. 7, we show the corresponding results with larger interleaver lengths from 1024 to 4096. All 3GPP LTE interleaver lengths have been considered, i.e., the interleaver lengths $1024, 1056, \ldots, 2016, 2048, 2112, \ldots, 4096$. In the search, all irreducible QPPs have been considered for all lengths. For comparison, we also show the results with optimized DRP interleavers up to length $N = 2560$ taken from [9, 23] (only estimates of the $d_{\min}$ are provided in [9, 23]) and the results with the 3GPP LTE QPPs from Table IV.

In Fig. 8, we compare the minimum distance for general optimum QPPs with the minimum distance for optimum QPPs with a quadratic inverse for the 3GPP LTE lengths from 736 to 4096 for 8-state turbo codes. From Fig. 8, we observe that the class of QPP-based interleavers with a quadratic inverse is in general inferior to the general class of QPP-based interleavers for larger block sizes. We remark that the interleaver length $N = 736$ is the first length among the 3GPP LTE lengths where the class of QPPs with a quadratic inverse is inferior to the general class of QPPs. For some lengths, the performance in the error floor region can be greatly improved by considering QPPs with no quadratic inverse. For instance, the optimum $d_{\min}$ increases from 38 to 50 for the lengths $\{1376, 1504, 1632, 1696, 1760, 1824, 1888, 1952\}$. Note that among the lengths giving an optimum $d_{\min}$ of 38 within the class of QPPs with a quadratic inverse, only the lengths $360, 432, 512, 896, 1008, 1344,$ and $1792$ do *not* satisfy the condition in (13) with $l = 0$. There are in total 42 lengths that give an optimum $d_{\min}$ of 38 within the class of QPPs with a quadratic inverse (when $N \leq 4096$).

Finally, we remark that a $d_{\min}$-optimum QPP interleaver is not necessarily unique, in the sense that there will be several QPPs with the optimum minimum distance. In Table V, we list some $d_{\min}$-optimum QPPs giving a $d_{\min}$ of at most 50 for some specific short-to-medium block lengths. For each QPP we also list the exact number $N_{d_{\min}}$ of minimum-weight codewords. We remark that these polynomials do not necessarily give the lowest multiplicity, but are also selected based on error rate performance through simulations.

### A. Improved QPPs

In this subsection, we present some improved QPPs (with respect to the induced turbo code $d_{\min}$) compared to the ones selected for the 3GPP LTE standard.

*Example 3:* For $N = 5504 = 2^7 \times 43$, the $d_{\min}$ is at most 55 over the entire class of QPPs. This follows from an exhaustive search over all QPPs with an inverse degree of at least four (see Example 2). Furthermore, the polynomial $f(x) = 21x + 1118x^2$ has an estimated $d_{\min}$ of 55 with an estimated multiplicity of 507, using the triple impulse method.

TABLE V
SOME $d_{\min}$-OPTIMUM QPPs FOR SOME PARTICULAR INTERLEAVER LENGTHS.

| $N$ | $f(x)$ | $f^{-1}(x)$ | $d_{\min}$ | $N_{d_{\min}}$ |
|---|---|---|---|---|
| 640 | $141x + 120x^2$ | $581x + 360x^2$ | 39 | 74 |
| 768 | $25x + 240x^2$ | $553x + 144x^2$ | 39 | 90 |
| 1024 | $245x + 448x^2$ | $861x + 832x^2$ | 45 | 486 |
| 1504 | $49x + 658x^2$ | $353x + 470x^2 + 1128x^3$ | 50 | 3241 |
| 2048 | $21x + 128x^2$ | $1853x + 1408x^2$ | 50 | 9198 |

TABLE VI
ESTIMATED/EXACT MINIMUM DISTANCE AND ITS CORRESPONDING ESTIMATED/EXACT MULTIPLICITY FOR SOME IMPROVED QPPs WITH A POLYNOMIAL INVERSE OF DEGREE GREATER THAN TWO. EXACT RESULTS APPEAR IN BOLD.

| $N$ | $f(x)$ | $f^{-1}(x)$ | $d_{\min}$ | $N_{d_{\min}}$ |
|---|---|---|---|---|
| 2496[a] | $119x + 702x^2$ | $215x + 390x^2 + 2184x^3$ | **51** | **3034** |
| 2624[a] | $125x + 1066x^2$ | $677x + 246x^2 + 2296x^3$ | **51** | **2556** |
| 2752[a] | $21x + 430x^2$ | $557x + 1634x^2 + 344x^3$ | **51** | **2853** |
| 2880[a] | $133x + 450x^2$ | $157x + 270x^2 + 1080x^3$ | **51** | **2986** |
| 2944[a] | $21x + 1196x^2$ | $701x + 1380x^2 + 2208x^3$ | **51** | **3236** |
| 3008[a] | $143x + 94x^2$ | $1951x + 470x^2 + 1128x^3$ | **51** | **2940** |
| 3200[a] | $83x + 100x^2$ | $347x + 1300x^2 + 2400x^3$ | **51** | **3132** |
| 3264[a] | $55x + 102x^2$ | $2359x + 2958x^2 + 2856x^3$ | **51** | **3396** |
| 3392[a] | $81x + 106x^2$ | $513x + 2862x^2 + 2968x^3$ | **51** | **3324** |
| 3456[a] | $91x + 108x^2$ | $1747x + 540x^2 + 864x^3$ | **51** | **3813** |
| 3520[a] | $27x + 110x^2$ | $1923x + 2750x^2 + 3080x^3$ | **51** | **3452** |
| 3648[a] | $43x + 114x^2$ | $403x + 1026x^2 + 2280x^3$ | **51** | **3580** |
| 3712[a] | $55x + 116x^2$ | $135x + 2900x^2 + 2784x^3$ | **51** | **3644** |
| 3776[a] | $1359x + 826x^2$ | $1151x + 354x^2 + 1416x^3$ | **51** | **3708** |
| 3904[a] | $1283x + 854x^2$ | $2715x + 2806x^2 + 2440x^3$ | **51** | **3836** |
| 3968[a] | $109x + 1054x^2$ | $1893x + 3162x^2 + 3720x^3 + 3720x^4$ | **52** | **242** |
| 4736[a] | $61x + 666x^2$ | $1941x + 814x^2 + 1480x^3 + 4440x^4$ | **53** | **145** |
| 5248[b] | $21x + 1886x^2$ | $2749x + 1066x^2 + 328x^3 + 328x^4$ | 53 | 161 |
| 5504[b] | $21x + 1118x^2$ | $2621x + 5418x^2 + 3784x^3 + 3784x^4$ | 55 | 507 |
| 6016[b] | $59x + 658x^2$ | $3059x + 4794x^2 + 3384x^3 + 5640x^4$ | 57 | 744 |
| 6144[c] | $59x + 1680x^2$ | $2291x + 4560x^2 + 1536x^3$ | **51** | **94** |

[a] The tabulated QPP is $d_{\min}$-optimal for this length.
[b] The tabulated QPP is $d_{\min}$-optimal for this length if the estimated $d_{\min}$ given in the fourth column is the exact value. In any case, the estimated $d_{\min}$ in the fourth column is an upper bound on the best possible $d_{\min}$ with QPP interleavers for this length.
[c] The tabulated QPP is $d_{\min}$-optimal for this length within the class of QPPs with an inverse degree of at most three.

*Example 4:* For $N = 6016 = 2^7 \times 47$, the $d_{\min}$ is at most 57 over the entire class of QPPs. This follows from an exhaustive search over all QPPs with an inverse degree of at least four; it follows from Proposition 3, the proof of Theorem 2, and Theorem 3 that it is sufficient to look at QPPs with an inverse degree of at least four. Furthermore, the polynomial $f(x) = 59x + 658x^2$ has an estimated $d_{\min}$ of 57 with an estimated multiplicity of 744, using the triple impulse method. The QPP from the 3GPP LTE standard gives a $d_{\min}$ of 51 with a much higher multiplicity of 6012.

In Table VI, we give some new QPPs for some 3GPP LTE interleaver lengths for which the estimated/exact $d_{\min}$ is strictly larger than 50. Also, the corresponding estimated/exact multiplicity $N_{d_{\min}}$ is given. The tabulated results (the $d_{\min}$ and the corresponding multiplicity) are exact when they appear in bold. In the third column of the table, an inverse polynomial, denoted by $f^{-1}(x)$, is tabulated. Note that for $N = 6144$, we have found a QPP giving a $d_{\min}$ of 51 with a very low multiplicity of 94, compared to the high multiplicity of 12179 for the 3GPP LTE turbo code. Also, for this length, the $d_{\min}$ is 50 for the 3GPP LTE turbo code. Finally, we remark that the highest found $d_{\min}$ is 51 when the QPP has an inverse polynomial of degree three.

## VI. Conclusion and Future Work

In this work, we have presented several upper bounds on the best achievable minimum distance with conventional binary turbo codes (with tailbiting termination, or dual termination when the interleaver length is sufficiently large) using QPP-based interleavers. We have verified by exhaustive computer search that the upper bounds are tight for larger interleaver lengths. One of the main results was a general upper bound of $2(2^{\nu+1}+9)$ on the minimum distance, independent of the interleaver length, when the QPP has a QPP inverse, where $\nu$ is the degree of the primitive feedback and monic feedforward polynomials. Furthermore, by means of several examples, we showed that by allowing the QPP to have a larger degree inverse may give strictly larger minimum distances (and lower multiplicities). We also provided the *exact* minimum distance and the corresponding multiplicity for all 3GPP LTE turbo codes (with dual termination). Finally, we computed the best achievable minimum distance with QPP interleavers for the 3GPP LTE lengths $N \leq 4096$ and compared the minimum distance with the one we get when using the 3GPP LTE polynomials.

Deriving a general, i.e., independent of $N$ upper bound on the $d_{\min}$ that holds for any interleaver length $N$ without any constraints on the degree of an inverse polynomial is an important open problem. Even finding a general upper bound when an inverse polynomial has degree three would be interesting. As a final remark in this respect, we identified the polynomial $39x + 760x^2 \pmod{9728}$ (inverse degree of three) with an estimated $d_{\min}$ of 56, which again indicates that 51 is probably *not* a universal upper bound on the $d_{\min}$ when the QPP has an inverse degree of three.